# Detector requirements for single mask edge illumination x-ray phase contrast imaging applications

G.K. Kallon,[a*] F. Vittoria,[a] M. Endrizzi,[a] P.C. Diemoz,[a] C.K. Hagen,[a] A. Zamir,[a] D. Basta,[a] and A. Olivo[a]

[a] *Department of Medical Physics and Biomedical engineering, University College London, Malet Place, Gower Street, WC1E 6BT London, United Kingdom*

E-mail: `gibril.kallon.10@ucl.ac.uk`

ABSTRACT: Edge illumination (EI) is a non-interferometric X-ray phase contrast imaging (XPCI) method that has been successfully implemented with conventional polychromatic sources, thanks to its relaxed coherence requirements. Like other XPCI methods, EI enables the retrieval of absorption, refraction and ultra-small angle X-ray scattering (USAXS) signals. However, current retrieval algorithms require three input frames, which have so far been acquired under as many different illumination conditions, in separate exposures. These illumination conditions can be achieved by deliberately misaligning the set-up in different ways. Each one of these misaligned configurations can then be used to record frames containing a mixture of the absorption, refraction and scattering signals. However, this acquisition scheme involves lengthy exposure times, which can also introduce errors to the retrieved signals. Such errors have, so far, been mitigated by careful image acquisition and analysis. However, further reduction to image acquisition time and errors due to sample mask/sample movement can increase the advantages offered by the EI technique, and enable targeting more challenging applications. In this paper, we describe two simplified set-ups that exploit state-of-the-art detector technologies to achieve single-shot multi-modal imaging.

---

[*] Corresponding author.

# Contents

## 1. Introduction

In addition to conventional X-ray absorption, X-ray phase contrast imaging (XPCI) methods enable the detection of phase contrast signals. Phase signals arise thanks to the unit decrement of the real part of an object's complex refractive index [1], and phase contrast is provided by the difference in this quantity between two media. Hence, weakly absorbing materials, which may be invisible in conventional absorption images can potentially be detected by XPCI set-ups. Some XPCI methods utilise optical elements to translate the phase modulations that are introduced by a sample into intensity variations, which can then be detected at some distance downstream from the object [2]–[6]. Among these techniques, we focus here on the edge illumination (EI) method, which has shown potential for translation to "real-world" applications [7], [8]. The EI method can be applied to both synchrotron and laboratory sources, thanks to its relaxed coherence and alignment requirements, and achromatic nature [7], [9]–[11].

## 2. Methods

### 2.1 Double mask edge illumination

The standard laboratory EI set-up employs two periodic masks with alternating transparent and opaque vertical apertures. The first mask is placed before the sample and is used to split a divergent cone beam into multiple, smaller beamlets. The second mask is placed before the detector to act as an analyser. Beamlets are aligned to straddle the opaque edges of the second mask, i.e. to partially impinge on the pixel. This is the "EI condition", which can be fulfilled by misaligning mask apertures with respect to each other. Figure 1 shows a schematic top-down view of the standard double-mask EI (DM-EI) set-up.

Frames that are recorded at a particular sample mask misalignment contain a mixture of absorption, refraction and ultra-small angle scattering (USAXS) signals. The proportion of each signal in the frame depends on the displacement of the sample mask, $\Delta x$, with respect to the position where the apertures of the two masks are aligned [12]; hence, the intensity, $I(\Delta x)$, measured at a given misalignment by a single pixel can be described as follows:

$$I(\Delta x) = A(L_{IC} * S)(x - R), \qquad (1)$$

where $L_{IC}$ represents the illumination curve (IC): a function that describes how the measured intensity changes as a function of mask displacement. The IC is used to characterise EI set-ups and is therefore measured without the sample, by scanning the first mask with respect to the second and recording the intensity at each scanning step. Equation (1) describes how the measured intensity can be modified by a sample at each misalignment. In particular, three signals are modelled: absorption ($A$) reduces the area beneath the IC curve, refraction ($R$) shifts its centre, and USAXS is represented by a convolution (indicated by *) between the IC and the object scattering function ($S$), which broadens the IC [12].

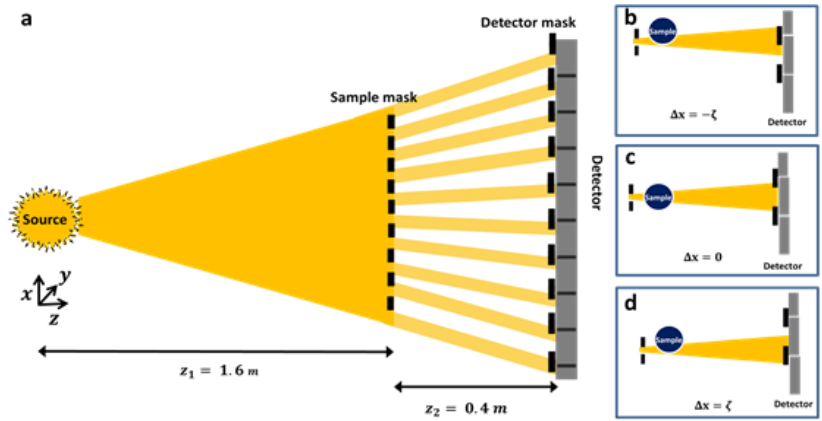

*Figure 1(a) A schematic top-down view of the standard, double mask edge illumination set up used to acquire the illumination curve. Panels (b)-(d) show, for a single detector pixel, three mask misalignments which are used to acquire mixed intensity frames of a sample.*

The second mask in the double mask EI (DM-EI) set-up (shown in figure 1(a)) fulfils two roles. First, it reduces the amount of background photons impinging on the detector; these photons contribute to the noise but not the signal, and therefore reduce the signal to noise ratio (SNR) of the data measured by the set-up. Second, it creates the edge needed for phase sensitivity by redefining the detector point response function (PSF), which also reduces the smoothing introduced by non-ideal detector PSFs. Hence, the second mask enables DM-EI set-ups to be implemented with various detectors while maintaining a relatively consistent performance.

However, the second mask also makes the set-up less dose efficient, since to separate the absorption, refraction and USAXS signals, three mixed intensity frames must be acquired under as many different illumination conditions in three separate exposures. Figures 1(b)-(d) show typical misalignments that can be used to acquire the three frames, which correspond to $\Delta x = -\zeta, 0, \zeta$. Note that the refraction signals recorded in the two symmetric positions $\Delta x = -\zeta$ & $\Delta x = \zeta$ will be inverted with respect to each other, while they are typically negligible at $\Delta x = 0$. Ultimately, this multi-shot approach makes DM-EI set-ups susceptible to positioning errors, which can arise when displacing the sample mask or sample between each frame. In addition, acquiring multiple exposures also increases the acquisition time and the dose delivered to a sample. In other words, it can present obstacles to the use of EI in "real-world" applications that require reduced doses and shorter acquisition times. Currently, errors arising from the sample and mask movements have been managed by careful image acquisition and analysis. However, these errors can be further reduced by exploiting recent improvements in detector technology.

In general, state-of-the-art detectors enable the construction of simplified XPCI set-ups that could not have otherwise been realised when the first laboratory implementations were developed. The new set-ups can perform single-shot multi-modal imaging, and therefore provide low-dose alternative implementations to multi-shot approaches. Recent research efforts have focused on developing single-shot multi-modal methods, both at synchrotrons and in the laboratory [13]–[19]. These developments have been motivated by the relative ease with which such methods can be adapted to more challenging applications compared to multi-shot approaches. For EI, a number of single-shot and single-scan imaging methods have been developed [20]–[23]. Among them, single-mask (SM) set-ups enable multi-modal imaging without making a priori assumptions on the sample composition, or requiring sample scanning for all applications. SM set-ups are constructed by eliminating the second mask and using only a single optical element. In this configuration, two or more frames can be acquired simultaneously under as many different illumination conditions [3], [22], [23].

Like DM-EI, EI based SM set-ups use a sample mask to split an incoming X-ray beam into multiple beamlets, which are aligned with the detector pixels. The detector pixels are then used to directly 'sense' or 'track' the changes suffered by the beamlets as a result of their interaction with a sample. We report here on two SM laboratory set-ups, based on the principles of 'sensing' and 'tracking' the beamlets, respectively, and show some of their preliminary results.

**2.2 Single mask edge illumination**

The first SM set-up can be referred to as the single-mask EI (SM-EI) set-up. SM-EI requires a sample mask with a period that is twice the pixel pitch, i.e. the mask is skipped to illuminate every other pixel when projected unto the detector [23]. The use of a 'skipped' mask allows the set-up to create an 'alternating illumination' condition, whereby each beamlet is aligned to sit at the boundary between two adjacent pixels. Pixel edges can then be used to directly 'sense' refraction-induced beam displacements, which cause the beamlet centre to be refracted towards one pixel, and away from its neighbour.

Figure 2 shows a top-down schematic of the SM-EI set-up, and can be used to demonstrate how the alternating illumination condition enables simultaneous acquisition of two images. Consider refraction that causes the first beamlet (at the top of the figure 2), which sits between pixels 1 and 2, to be displaced upwards, towards pixel 1. This beamlet displacement would increase the counts recorded by pixel 1, while simultaneously reducing those measured by pixel 2. Effectively, each set of skipped pixels (i.e. even and odd ones) 'sees' an inverted refraction signal with respect to its neighbour, but the same absorption signal. These sets of pixels can therefore be separated and used to create two mixed intensity frames with inverted refraction signals but the same absorption signal.

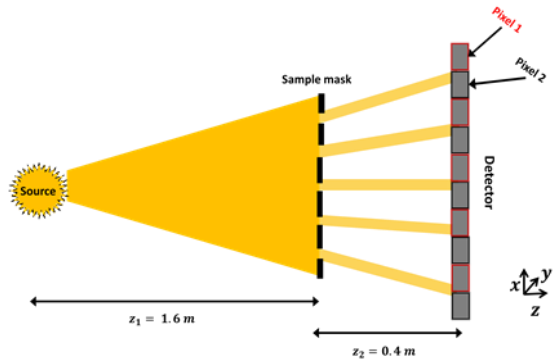

*Figure 2 A schematic of the single mask edge illumination set-up.*

Two of the three previously mentioned signals can be retrieved using the SM-EI set-up. For weakly scattering samples, the absorption and refraction signals can be retrieved by neglecting the scattering term in equation (1). On the other hand, for weakly refracting samples, an alternative alignment can be used to obtain the absorption and USAXS signals [24].

However, the use of a line skipped mask means the sampling rate of the SM-EI set-up is halved with respect to the DM-EI set-up. This can be a reasonable trade-off for some applications where a high sampling rate is not required [8]. It must also be noted that the SM-EI set-up has, thus far, only been realised using direct conversion detectors, which typically have pixels sizes $\geq 50\ \mu m$ and possess limited pixel cross-talk. These conditions significantly improve the benefits offered by the SM-EI set-up.

The SM-EI set-up was used to acquire mixed intensity frames of a number of wires (in air). A Rigaku 00HF source with a molybdenum target was operated at 35 kVp and 25 mA, and the Anrad SMAM amorphous selenium flat panel detector was used, with a pixel size of $85\ \mu m$. The sample mask, which has a period of $134\ \mu m$ and an aperture size of $16\ \mu m$, and the detector, were placed at $1.6\ m$ and $2\ m$ away from the source, respectively. These distances ensured that the alternating illumination condition was fulfilled, i.e. the magnified sample mask period was equal to twice the detector pitch, i.e. $170\ \mu m$.

**2.3 Beam tracking set-up**

The "beam-tracking" (BT) set-up is a single-mask technique that can be realised by using a high resolution detector or a high magnification set-up to resolve the footprint of beamlets falling over three or more pixels. The beamlet spatial intensity distribution, before and after its interaction with the sample, are compared against each other to retrieve the absorption, refraction and scattering signals in a single shot [22].

Equation (1) can also be used to model the intensity measured by the BT set-up; however, the $L_{IC}$ term, which represents the IC for DM-EI is replaced by a term that represents the intensity distribution of the individual beamlets, $L_{BT}$. Thus, absorption is modelled as a reduction in the beamlet amplitude; refraction, as a shift of its centre, and scattering as its broadening. A schematic of the BT set-up is shown in Figure 3.

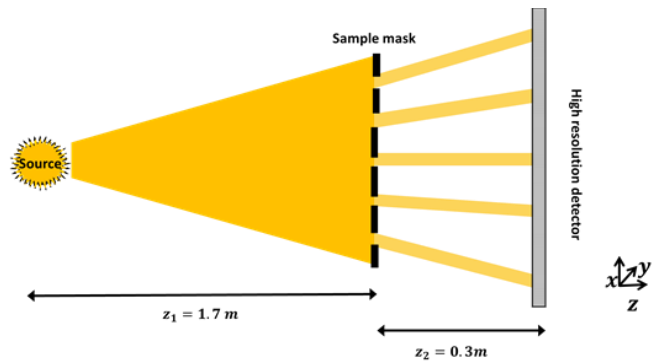

*Figure 3 A schematic of the beam-tracking set-up.*

The BT set-up has been used at both synchrotrons and laboratories, for planar and computed tomography (CT) imaging [22], [25]. BT set-ups have so far only been realised using high resolution detectors, which typically possess small fields of view, or using high magnification set-ups with inefficient indirect conversion detectors.

For this work, the BT set-up was constructed using a rotating tungsten source, operated at 46 kVp and 26 mA, and a high resolution CCD camera coupled through a fibre optic plate to a Gadox scintillator, with an effective pixel size of 4.54 µ$m$ (Photonic Science). The mask had a period of 93 $\mu m$ and an aperture size of 9 $\mu m$.

## 3. Results

The intrinsic spatial resolution of these set-ups can be well approximated by the sample mask aperture size, while object sampling is determined by the sample mask period. Finer object sampling can be achieved through "dithering", which involves displacing the sample in multiple sub-pixel steps, and acquiring a frame at each step. Individual dithering steps can then be stitched together to obtain a higher resolution image. Dithering was used to acquire both sets of data presented here; 32 dithering steps were used for the SM-EI set-up, and 10 dithering steps were used for the BT set-up. Evidently, when dithering is used, neither SM set-ups remain a single-shot technique. However, as previously mentioned, for some applications dithering may not be necessary, in which case both techniques maintain this advantages—this is dictated uniquely by the required resolution.

For both set-ups, the mixed intensity frames are not shown, we will show instead the resulting retrieved images. Refraction and transmission planar images will be shown for the SM-EI set-up, while CT projections of absorption, phase and USAXS will be displayed for the BT case. Note that the phase image is obtained by performing a line integral over the refraction image, hence it is not differential in nature (does not contain the characteristic dark & bright fringes of refraction images, like the one shown in figure 4a)

### 3.1 Wire profiles extracted from experimental measurements

For the SM-EI set-up, a large number of dithering steps was used to ensure that the refraction peaks of the wire were finely sampled. Figures 4(a) and (c) show the refraction and transmission images of the four wires, which were retrieved using the phase retrieval algorithm described by Diemoz et al 2013 [26].

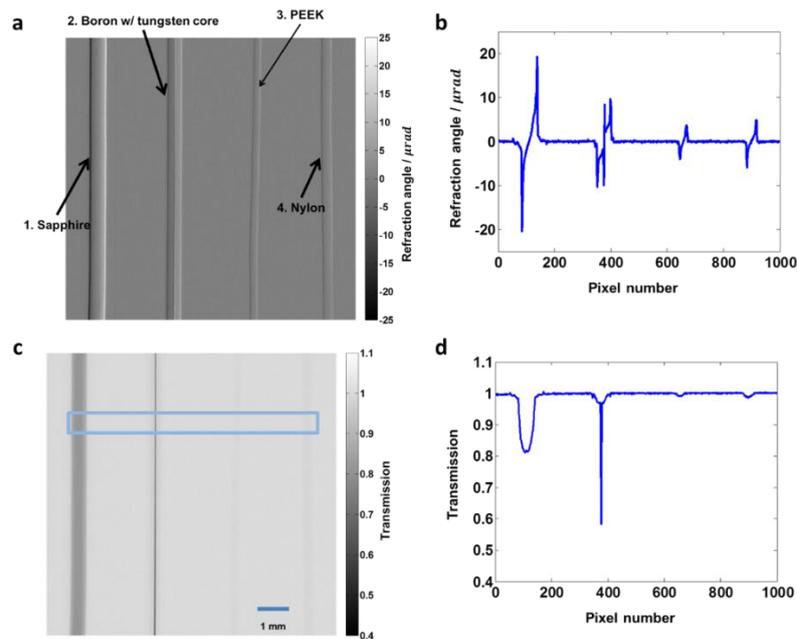

*Figure 4 (a) Retrieved refraction and (c) transmission images obtained from experimental measurements with the SM-EI set-up. The four wires are in the phantom are labelled in (a) as 1. sapphire, 2. Boron with a tungsten core, 3. PEEK and 4. Nylon. Panels (b) and (d) show the refraction and transmission profiles, respectively, which have been extracted as an average over multiple lines from the ROI shown in (c).*

All the wires are clearly visible in the refraction profile shown in figure 4(b), thanks to phase contrast effects, which manifest strongly at object boundaries, and can be seen as bright and dark fringes in figure 4(a). However, PEEK and Nylon wires, which are weakly absorbing, are only barely visible in the attenuation profile in figure 4 (d) and its corresponding image. On

– 6 –

the other hand, the sapphire wire and the tungsten core, which are both strongly absorbing and refracting materials, are both clearly visible in figures 4(a) and (c).

### 3.2 Beam-tracking computed tomography

For the BT set-up, CT images of a toothpick were acquired using 1441 angular projections each with 10 sub-pixel dithering steps; therefore, the scanning step size was equal to $9.5\ \mu m$. The signals extracted from each step are then recombined in a single "oversampled" image. This process was repeated for different angular position of the sample in the angular range $[0°, 180°]$ to obtain a complete tomographic scan. Detector dark current was measured by acquiring 10 frames before and after each scan with the source turned off, these were then averaged and subtracted from all the other measurements. In addition, flat field images, i.e. images without the sample, were acquired during the scan, after every 20 angular positions.

Before performing the CT reconstruction, the retrieved refraction signal was numerically integrated along the x direction. CT reconstructions were performed with an iterative reconstruction algorithm based on a Gradient Descent minimization approach with Tikhonov-Miller regularization; however, the regularization was not applied to the phase image, thanks to its low noise level. The reconstruction algorithm requires an initial guess of the reconstructed slice and the experimentally measured sinogram as inputs.

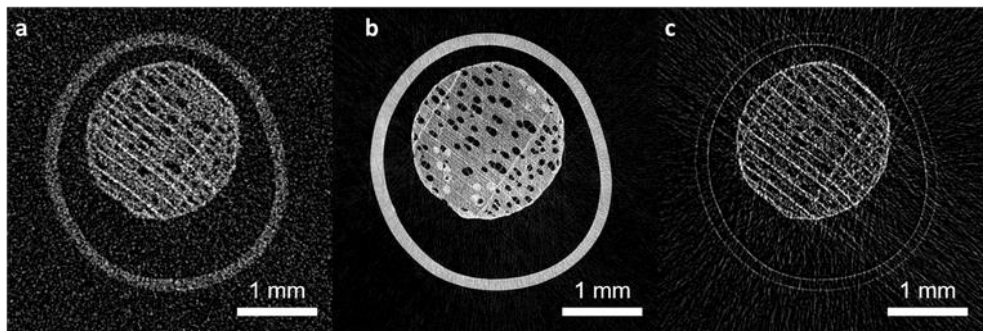

*Figure 5 A transverse cross-section of a toothpick placed inside a cylindrical container for (a) absorption, (b) phase and (c) USAXS channels.*

Figures 5 (a)-(c) show the absorption, phase and USAXS signals, respectively. The absorption signal is weak, as expected for this sample. Conversely, in figure 5(b), the phase image shows strong contrast for the larger internal structures of the toothpick. Finally, figure 5(c) shows the USAXS signal, which originates from structural inhomogeneities within the sample, i.e. structures that are too small to be resolved by the pixel. This particular signal channel is particularly susceptible to image noise, the prevalence of which explains the reduced quality of the image. However, the presence of sub-pixel structures is clearly highlighted by the brightest parts of the image.

## 4. Conclusion

We have described two single mask XPCI techniques (SM-EI and BT) that enable single-shot multi-modal imaging. We have also discussed some advantages and disadvantages of both SM set-ups. We showed planar refraction and absorption images of four wires that were acquired with the SM-EI set-up, as well as CT projections of the absorption, phase and USAXS signals of a toothpick acquired with the BT set-up. SM-EI set-ups have the potential to address some limitations of the DM-EI set-up. However, the new implementations rely on the improvements offered by state-of-the-art detectors, i.e. sharper, more ideal PSFs, high efficiency, high resolution detectors etc. SM-EI and BT set-ups are less complex, alternative implementations of XPCI imaging. They use one optical element, making them easier to align. This simplicity also makes them more robust against vibrations and subsequent misalignments during the image acquisition procedure. Furthermore, SM-EI set-ups provide a means of reducing the acquisition time and dose delivered to the sample, which can be important in for "real-world" applications.

## Acknowledgments

This work was supported by the EPSRC grants (EP/N509577/1, EP/M507970/1 and EP/1021884/1). ME was supported by the Royal Academy of Engineering under the RAEng Research Fellowships scheme. PCD is supported by the Marie Curie Career Integration Grant PCIG12-GA-2012-333990 within the Seventh Framework Programme of the European Union.